# A model of an optical biosensor detecting environment

Anh D. Phan, Dustin A. Tracy
Department of Physics, University of South Florida
Tampa, Florida 33620, USA
anhphan@mail.usf.edu

N. A. Viet
Institute of Physics
Hanoi 10000, Vietnam
vieta@iop.vast.ac.vn

*Abstract*—**Heller et. Al. (Science 311, 508 (2006)) demonstrated the first DNA-CN optical sensor by wrapping a piece of double-stranded DNA around the surface of single-walled carbon nanotubes (CN). This new type of optical device can be placed inside living cells and detect trace amounts of harmful contaminants by means of near infrared light. Using a simple exciton theory in nanostructures and the phenomena of B-Z structural phase transition of DNA, we investigate the working principle of this new class of optical biosensor from DNA by using the nanostructure surface as a sensor to detect the property change of DNA as it responds to the presence of target ions. We also propose some new design models by replacing carbon nanotubes with graphene ribbon semiconductors.**

## I. INTRODUCTION

Explanations of many biological processes and the development of some new applications require an exact description of DNA interactions with various nanostructures. Scientists from a wide range of fields recently employed DNA as a potential material for sensitive biosensor designs [1-3]. The complexity of DNA and DNA combining systems provide a variety of significant challenges to the development of such biosensors.

The physical properties of DNA vary with temperature and solution conditions [4]; therefore, these conditions need to be managed to accurately control biodevices. DNA exists in three main conformations, the A, B, and Z-forms. Much attention is given to the B-Z transition of DNA because it possesses a unique feature that can be utilized for environmental change detection. The surrounding medium significantly effects DNA conformation determination [4-6]. Because the DNA structure is sensitive to changes in its environment, DNA is a promising material for the construction of biosensors.

Carbon nanotube (CNT) and graphene nanoribbon (GNR) based sensors are of major interest due to their tremendous promise for obtaining sequence specific information in a faster, simpler and cheaper manner than traditional analysis [6,8-10]. These two carbon systems provide a number of treatment based and sensor based applications in the liquid and wastewater treatment industries. The unique assembly properties of DNA together with the unusual optical characteristics of graphene nanoribbons and carbon nanotubes suggest that DNA based biosensor may become key devices in the near future.

We employ a simple model of DNA wrapping CNT and GNR in order to theoretically investigate the effective dielectric function of a medium. The model is in good agreement with experiments from previous studies [6,7]. The dielectric function strongly relates to external factors, including temperature and concentration of ions. These factors, therefore, significantly influence the exciton binding energy of DNA-carbon-system based sensors and suggest that the biosensors can be detectors for environmental changes with or without the conformational transformations of DNA.

The rest of the paper is organized as follows: In Sec. II, the theoretical structure model of DNA is introduced in order to calculate the effective dielectric constant. In Sec. III, the change of the dielectric due to temperature and ionic concentration is given. The exciton binding energy of the whole system and the principle of the biosensor are discussed in Sec. IV.

## II. THEORETICAL MODEL OF DNA

The double stranded DNA coil has a helical configuration. Our biosensor models in Fig.1 are based on DNA's ability to wrap around other nanostructures. Characteristic lengths are the radius $r$, the pitch $b$ along the axis of the helical DNA, and the width $a$ of a DNA strand.

In this model, the effective dielectric constant of DNA and its surrounding medium is expressed as [6]

$$\varepsilon = f\varepsilon_{DNA} + (1-f)\varepsilon_W, \quad (1)$$

where $\varepsilon_{DNA}$ and $\varepsilon_W$ are the dielectric constants of the DNA and the solution, respectively, and $f$ is the ratio of the DNA-covered surface area to the total surface area of the nanostructures. The model, therefore, suggests that $f$ is dependent on the solvent geometries. In our calculations, we use CNT and GNR to be components of the biodevice.

In the case of CNT wrapped by DNA, $f$ is expressed by [6]

$$f = \frac{a}{r_0 + r}\sqrt{\frac{b_0^2 + r_0^2}{b_0^2 + r_0^2 - (r_0+r)r_0}}, \quad (2)$$

When DNA wraps GNR, the expression of the ratio $f$ is given by [11]

$$f = \frac{a}{2W}\sqrt{\frac{4\pi^2 r_0^2 + b_0^2}{4\pi^2 r_0^2 + b_0^2 - 4\pi r_0 W}}, \quad (3)$$

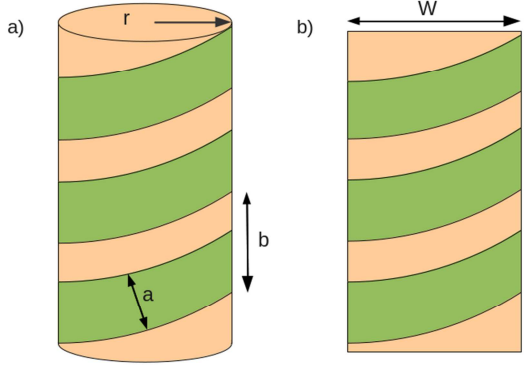

Figure 1. (Color online) The schemes of a) CNT-DNA based biosensor model and b) GNR-DNA based biosensor model.

here $r_0, b_0$ and $a$ are 1, 3.32 and 0.51 nm, respectively, for B DNA; and $r_0 = 0.9$ nm, $b_0 = 4.56$ nm, and $a = 1.18$ nm for Z DNA [6], $r$ is the nanotube radius and $W$ is the width of GNR.

### III. DIELECTRIC FUNCTION OF SYSTEMS

In real biological systems, the dielectric function of the solution around DNA is strongly influenced by solute ions such $Na^+$ or $Mg^{2+}$, which are abundant in DNA's natural cellular environment. The nature of DNA-ion interactions is complex because there are many competing factors, including chain configuration, ionic concentration, ionic size and ionic shape. The presence of the ions causes the B-Z transition of DNA [5]. The configuration of DNA has the B-form at low salt concentration and the Z-form at high salt concentration [11].

The dielectric function is also temperature dependent. A simple explanation for this occurrence relies on the abundance of water in the solution. Authors in Ref.[12] proved that dielectric constant of water decreases with a reduction in temperature. These results suggest that the dielectric function of the solution surrounding DNA can be divided into two parts corresponding to two variables [4], temperature and salt concentration, $\varepsilon_W \equiv \varepsilon_W(T, C) = \varepsilon_W(T)h(C)$. Here $\varepsilon_W(T)$ is the dielectric constant as a function of temperature $T$, and $h(C)$ is the salt concentration correction versus molarity $C$ of $Na^+$. The expressions of $\varepsilon_W(T)$ and $h(C)$ are given by

$$\varepsilon_W(T) = 249.4 - 0.788T + 7.2 \times 10^{-4}T^2, \quad (4)$$

$$h(C) = 1 - 0.255C + 5.15 \times 10^{-2}C^2 - 6.89 \times 10^{-3}C^3. \quad (5)$$

The results in Ref.[6,11] were calculated at room temperature and $C = 0$, so $\varepsilon_W = 80$. Although authors confirmed that the number of ions causes the B-Z transition, they did not consider the dielectric function change versus ionic concentration. In this case, the theoretical results are quite suitable for experimental measurements because the B-Z transition may occur at low concentration of $Hg^{2+}$. The dielectric function, therefore, can be independent of the $Hg^{2+}$ concentration. In our calculations, the concentration of ions must be considered due to the transformation from the B-form to the Z-form at $C_0 = 2.25$ M [4]. Specifically, the salt concentration directly affects the conformation transitions [13].

A consideration of the effect of temperature on the dielectric function gives some remarkable insight. Each type of pure DNA has an individual critical temperature $T_c$ [14]. An increase in temperature commonly leads to a stretching of DNA strands. When $T \geq T_c$, DNA begins to break. In our systems, the interaction between DNA and CNT or GNR is so large that this tearing of DNA as a function of temperature can be ignored. Thus, the dielectric constant of DNA remains unchanged; $\varepsilon_{DNA} = 4$ [6].

Fig. 2 shows the effective dielectric constants for CNT (8,7) and GNR with width 1.17 nm. Within the given range, the dielectric constant and its corresponding rate of change decrease with increases in temperature and concentration. Calculations performed on CNTs with radii between 0.379 nm and 0.516 nm and GNRs with widths between 0.43 nm and 2.87 nm show very similar trends, although the effective dielectric constant is found to increase slightly in response to increases in GNT radii and GNR width. The left-handed Z-DNA conformation is slightly thinner and longer than the right-handed B-DNA. This discrepancy causes the change in $f$ and the effective dielectric constant.

An increase in temperature creates more space between the molecules of the solutions and is a reason for a decrease in $\varepsilon$. Interestingly, the presence of ions reproduces the decrement of dielectric response. The sodium ions will replace the much larger water molecules. When kept at the same total molecular concentration, this replacement will increase free space. The electrostatic forces of ions in the liquid induce the structural change of solvent molecules causing a reduction in $\varepsilon$ [15,16]. This effect reproduces the hydration shells orienting molecules.

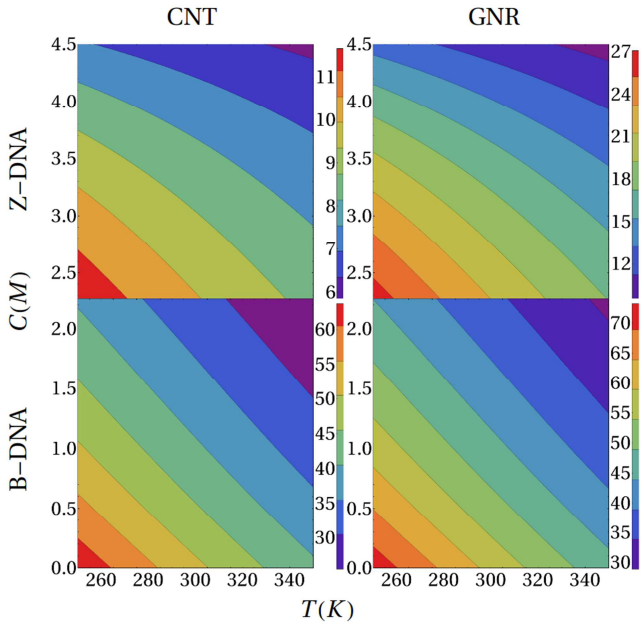

Figure 2. (Color online) The effective dielectric function as a function of temperature T (K) and concentration C (M).

## IV. EXCITONIC TRANSITION ENERGY OF BIOSENSORS

An analytical formula that allows for the calculation of the exciton binding energy $E_b$ of the DNA-CNT biosensor can be expressed by [6,17]

$$E_b = A_b r^{\alpha-2} \mu_c^{\alpha-1} \varepsilon^{-\alpha}; \quad (6)$$

$\mu_c$ is the reduced mass of exciton in CNT, $\alpha = 1.4$, and $A_b = 24.1\ eVnm^{3/5}$ [6] for CNTs that have the radius in the 0.5-1.25 $nm$ regime.

The armchair graphene nanoribbon (AGNR) is the most semiconductive GNR that has a large band gap; therefore, it should be chosen for clearly observing the change in the exciton binding energy of the DNA-GNR biosensor. The expression in Eq.(7) defines $E_b$ as [17]

$$E_b = A_f W^{\beta-2} \mu_G^{\beta-1} \varepsilon^{-\beta}, \quad (7)$$

where $\beta$ is 0.98 or 1 depending on the form of the dimer number 3p or 3p+1, respectively. $\mu_G$ is the reduced mass of the exciton in AGNR [17] and p is an integer.

In Fig.3 the excitonic energy difference $\Delta E_b = E_{bZNA} - E_{bBNA}$ at the B-Z transition deserves further consideration. The $\Delta E_b$ values are approximately 85, 52 and 35 meV for CNTs (6,5), (10,2) and (9,5), respectively. These values are larger than those values obtained when the biosensor is in contact with $Hg^{2+}$ as in Ref.[6]. This type of biosensor, thus, is more sensitive in NaCl salt than in $HgCl_2$.

A significant variation of $E_b$ does not exist while DNA is in the B-form. $E_b$ of B-DNA has a maximum shift of 5 meV while C varies by 0-2.25 M for these three CNTs. The biosensor, therefore, has difficulty detecting concentration changes in these conditions. $\varepsilon$ is much smaller in the Z-form DNA solution that it is in the B-DNA solution. The smaller $\varepsilon$ leads to a dramatic change in $E_b$ at $T = 300\ K$. Although $E_b$ is inversely proportional to the radii of CNTs, the exciton binding energy of the biosensors increases to 40 meV when C approaches 4.5 M. This sensor can be manufacture to detect the concentration in the environment or in living bodies above the critical concentration.

Both protein and DNA have been studied at low temperatures [18,19]. The water in the immediate vicinity of the molecule does not freeze at 273 K. Even at a temperature of 200 K, the water is not fully frozen. Moreover, changes in temperature cause little variation at the critical concentration [5]. We can, therefore, investigate the behavior of the biosensors in the temperature range of 250-350K with constant $C_0$.

The exciton binding energy of DNA-AGNR biosensor is plotted in Fig.4. The widths of the GNRs are 1.15, 1.54, and 2.27 nm, equivalent to the form of dimer number 3p. Previous calculations [11] showed that GNRs with the widths 3p provide the best effective biosensors compared to other types of GNR. Like Fig.3, $E_b$ is interrupted at the critical concentration for these three AGNR. Although $\Delta E_b$ in NaCl is larger than $\Delta E_b$ in $HgCl_2$ [11], it is smaller than $\Delta E_b$ of a DNA-CNT biosensor in the same medium. The findings demonstrate that the CNT based sensors are more sensitive than the AGNR based sensors.

Another feature is that $\Delta E_b$ is more sensitive to temperature in the DNA-AGNR than in the DNA-CNT biosensor. This result and comparison agree with the calculations in [11].

The DNA-AGNR biosensor has the important capability of detecting the change of ionic concentration in an environment. In the B-form, the exciton binding energy shifts approximately 12 meV in the concentration range of 0-2.25 M. Particularly, an exciton binding energy difference of 50 meV exists between the C value of 2.25 and 4.5 M. As temperature varies in the 250-350K regime, the discrepancies of $\Delta E_b$ are less than 6 meV. Meanwhile, the change of $\Delta E_b$ in the DNA-CNT biosensors in the same temperature range are larger than 10 meV but smaller than 20 meV. As a result, DNA-CNT biosensors can be used to detect temperature variations.

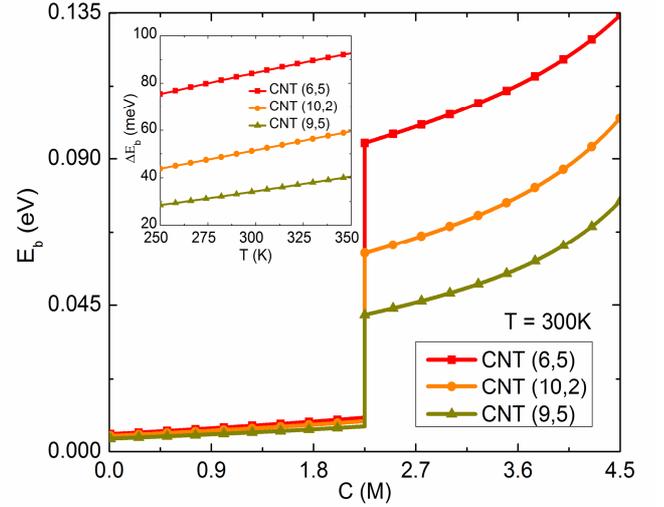

Figure 3. The exciton binding energy of DNA-CNT biosensor as a function of temperature T (K) and concentration C (M).

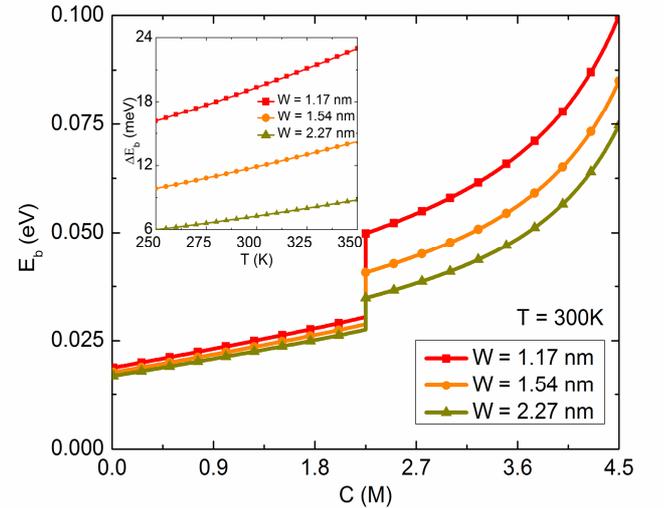

Figure 4. The exciton binding energy of DNA-AGNR biosensor as a function of temperature T (K) and concentration C (M).

GNRs can be created through the unzipping of CNTs. As a consequence, the sensors made with CNT or AGNR have the same behaviors in optical properties, particularly in exciton problems, as those made from GNR. Their exciton binding energies and band gaps are strongly dependent on the external field, including the electric field and magnetic field [20,21]. Currently, using CNT based sensor is more beneficial than using sensors manufactured by AGNR because of the simplicity of the CNT design. The radius of CNT can be controlled. The creation of sub-10-nm-GNR is a large barrier for experimentalists to overcome. Due to the fact that GNR possess unusual properties, such as the ability to become a half-metallic material when an in-plane external field is applied [22], GNR is a promising candidate for next generation devices.

For regular bulk systems, the parameters $\alpha$ and $\beta$ in Eq.(6) and (7) are equal to 2. In the case of GNR and CNT, the values are smaller because the screening of electrons plays a less important role in two dimensional systems [17]. The environment has a larger influence on the exciton binding energy of three dimensional systems than the presence of certain CNTs or GNRs.

ACKNOWLEDGMENT

The work was partly funded by the Nafosted Grant No. 103.06-2011.51.